\documentclass[11pt,twoside]{article}
\usepackage{asp2014}

\aspSuppressVolSlug
\resetcounters

\begin{document}

\title{Creating High Quality All-Sky Visualizations of Astronomy Image Data Sets: HiPS and Montage}

\author{G. Bruce~Berriman$^1$, John C. Good$^2$, Vandana Desai$^2$ and Steven L. Groom$^2$} 
\affil{$^1$Caltech/IPAC, Pasadena, CA~~  91125, USA; \email{gbb@ipac.caltech.edu}}
\affil{$^2$ Caltech/IPAC, Pasadena, CA~~  91125, USA} 

\paperauthor{Bruce~Berriman}{gbb@ipac.caltech.edu}{0000-0001-8388-534X}{Caltech/IPAC}{}{Pasadena}{CA}{91125}{USA}
\paperauthor{John~C.~Good}{jcg@ipac.caltech.edu}{ORCID}{Caltech/IPAC}{}{Pasadena}{CA}{91125}{USA}
\paperauthor{Vandana~Desai}{desai@ipac.caltech.edu}{0000-0002-1340-0543}{Caltech/IPAC}{}{Pasadena}{CA}{91125}{USA}
\paperauthor{Steve~Groom}{sgroom@ipac.caltech.edu}{0000-0001-5668-3507}{Caltech/IPAC}{}{Pasadena}{CA}{91125}{USA}



  
\begin{abstract}
We describe a case study to use the Montage image mosaic engine to create maps of the ALLWISE image data set in the Hierarchical Progressive Survey (HiPS) sky-tesselation scheme. Our approach demonstrates that Montage reveals the science content of infrared images in greater detail than has hitherto been possible in HiPS maps. The approach exploits two unique (to our knowledge) characteristics of the Montage image mosaic engine: background modeling to rectify the time variable image backgrounds to common levels; and an adaptive image stretch to present images for visualization. The creation of the maps is supported by the development of four new tools that when fully tested will become part of the Montage distribution. The compute intensive part of the processing lies in the reprojection of the images, and we show how we optimized the processing for efficient creation of mosaics that are used in turn to create maps in the HiPS tiling scheme. We plan to apply our methodology to infrared image data sets such a those delivered by Spitzer, 2MASS, IRAS and Planck.
\end{abstract}

\section{Introduction}
In a Hierarchical Progressive Survey (HiPS) map, progressive zooming will reveal an image sampled at ever smaller or larger spatial scales. HiPS was pioneered by the Centre de Donne\'es de Strasbourg (CDS) \citep{2015A&A...578A.114F}, and has subsequently been recommended as a standard by the International Virtual Observatory Alliance (IVOA). HiPS is based on the HEALPix sky tiling scheme, which subdivides the celestial sphere into levels of progressively smaller cells with the same surface area on the celestial sphere \citep{2005ApJ...622..759G}.

As of October 2019, there are 676 HiPS maps available through 21 servers. The maps were created with the CDS HiPSGEN application, intended for fast processing of large image data sets. As \citet{2015A&A...578A.114F} realized, HiPSGEN's use of a background level assumed constant across an image or derived from a group of pixels limits the quality of the maps in the infrared, where the background radiation is usually very high.  We describe here how the functionality of the  Montage image mosaic engine improve the quality of infrared HiPS maps \citep{2017PASP..129e8006B}. Montage supports HEALPix; rectifies backgrounds across images to a common level by modeling the differences in background levels between images; supports an adaptive image stretch algorithm; and offers easy parallelization.

\section{Creating HiPS maps with Montage}

 HiPS maps involve creating a base mosaic at the lowest required HEALPix level, usually chosen to match as closely as possible the spatial sampling of the input images, then cutting out the HiPS cells in PNG format from this mosaic. The process is repeated at successive HEALPix levels, per the prescription in the HiPS specification \footnote{\url{http://www.ivoa.net/documents/HiPS/20170406/PR-HIPS-1.0-20170406.pdf}}. We have developed four new modules, listed in Table {\ref{Table1}}, to support the creation of HiPs maps. The modules will be released as part of the Montage distribution when testing is complete.

\begin{table}[!ht]
\caption{New Montage Modules To Enable HiPS Creation}
\smallskip
\begin{center}
{\small
\begin{tabular}{ll}  
\tableline
\noalign{\smallskip}
Module & Functionality \\
\noalign{\smallskip}
\tableline
\noalign{\smallskip} 

mHPXMosaicScripts & Scripts to reproject the data into a set of plates \\
                  & at the lowest HEALPix level.\\
HiPSSetup         & Create directory tree for HiPS maps. \\
mHPXShrinkScripts & Build the plates in FITS format at all levels.  \\
mHiPSPNGs        & Make color PNG tiles from the FITS files. \\
\noalign{\smallskip}
\tableline
\end{tabular}
}
\end{center}
\label{Table1}
\end{table}
\noindent 

 {\tt mHPXMosaicScripts} creates the base mosaic from which the HiPS tiles will be created.  {\tt HiPSSetup} creates the directories in which the HiPS tiles will be stored. {\tt mHPXShrinkScripts}  populates the directories with HiPS compliant tiles in FITS format. It makes use of the Montage module {\tt mShrink}, according to the following recipe:  create a header describing the geometry of the HiPS tile;
 shrink the image; repeat for the next level. {\tt mHiPSPNGs} then processes the FITS images to create the HiPS tiles in PNG format, and then creates an all-sky histogram to determine the image stretch for display. These PNG files, with the stretch applied, are what is shown in HiPS viewers.

\section{Creating ALLWISE HiPS Images}
We have conducted a successful pilot project that created HiPS maps down to HEALPix level 18 (scale 0.8 arcseconds) of images acquired in the Galactic Plane by the ALLWISE survey \footnote{\url{http://wise2.ipac.caltech.edu/docs/release/allwise/}} in bands 1 (3.4 $\mu$m),3 (12$\mu$m), and 4 (22 $\mu$m). The ALLWISE image data set is large enough to require  processing at scale and the complex infrared backgrounds stress the background rectification algorithms. The processing used an  80-core cluster whose jobs were managed by Slurm, and exploited the fast reprojection algorithm  encoded in {\tt mProjectQL} (see below), to create a mosaic stored in 80 x 80 plates, with overlaps between them. The background rectification was performed across the images that compose each tile, and then across all the tiles.The three color map of the Galactic plane in Figure \ref{Figure 1} shows how Montage successfully rectified the background levels to reveal the science content of the images. 

Given the success of the approach, we have begun a production run with the Slurm cluster to deliver all-sky HiPS maps in all four ALLWISE bands. Figure \ref{Figure 2} shows one of the first images created in this run, in ALLWISE Band 1 (3.4 $\mu$m). The anticipated processing time for all bands on the 80-core cluster 4.8 days. This estimate is broken down by task as shown in Table \ref{Table2}. The table it makes clear that the very first step, reprojection, is the most time intensive task. This is despite the fact that we are using the fastest of three reprojection algorithms supported by Montage, based on Lanczos resampling and delivered in the {\tt mProjectQL} module. This reprojection is strictly not flux conserving, but is more than adequate for visualization.

\begin{table}[!ht]
\caption{Estimated processing times for  creation of the ALLWISE HiPS maps}
\smallskip
\begin{center}
{\small
\begin{tabular}{lr}  
\tableline
\noalign{\smallskip}
Process  & Time (s)\\
\noalign{\smallskip}
\tableline
\noalign{\smallskip}
Reprojection                        &   1800 \\
Background modeling/rectification   &  450 \\
Coaddition                          &   50\\
PNG generation                      &  220 \\
Total time per plate                & 2,520 \\
\noalign{\smallskip}
\tableline 
\noalign{\smallskip}
Estimated times:                    &       \\
Total time per band (80 cores)      &  1.2 d \\
Total time for 4 bands              & 4.8 d \\
\noalign{\smallskip}
\tableline
\end{tabular}
}
\end{center}
\label{Table2}
\end{table}
\noindent

\section{Future Plans}
Montage is a powerful tool for creating HiPS maps of infrared image data sets. The modest augmentations to functionality to support HiPS will be released presently. Future plans are on two broad fronts: deliver HiPS maps of important infrared imaging data sets such as those acquired with Spitzer, Planck, 2MASS, IRAS...; and investigate how cloud platforms can be employed to compute HiPS maps at scale.

\acknowledgements Montage is funded by the National Science Foundation under Grant 1835379. ALLWISE makes use of data from WISE, which is a joint project of the University of California, Los Angeles, and the Jet Propulsion Laboratory/California Institute of Technology, and NEOWISE, which is a project of the Jet Propulsion Laboratory/California Institute of Technology. WISE and NEOWISE are funded by the National Aeronautics and Space Administration. We thank Pierre Fernique and Thomas Boch for discussions on HiPS creation.

 \bibliography{O5-2}


\articlefigure {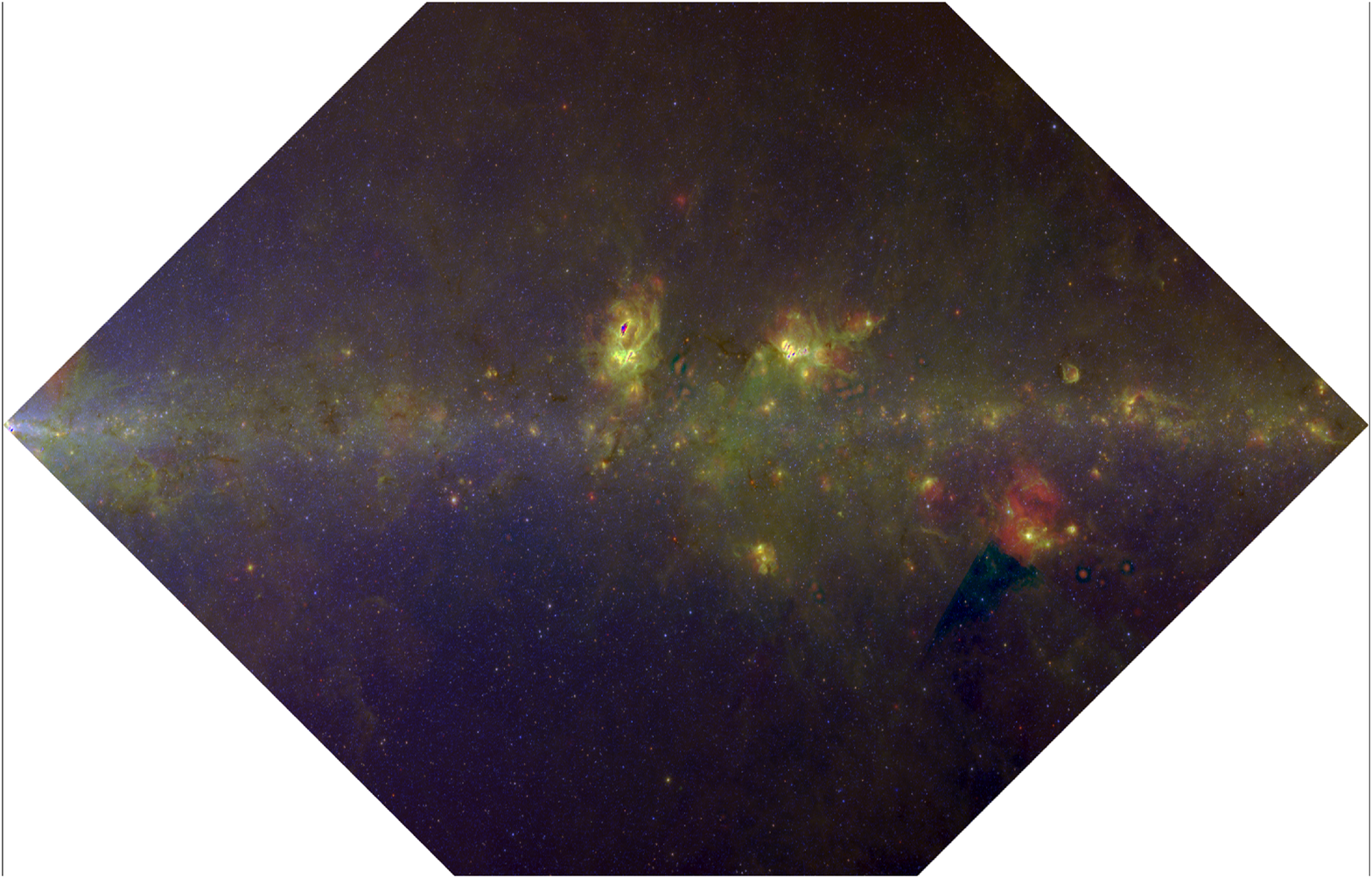}{Figure 1}{Three color HiPS map of a section of the Galactic Plane at level 4 computed with Montage for ALLWISE bands 1 (3.4 $\mu$m), 3 (12$\mu$m) and 4 (22 $\mu$m). }

\articlefigure{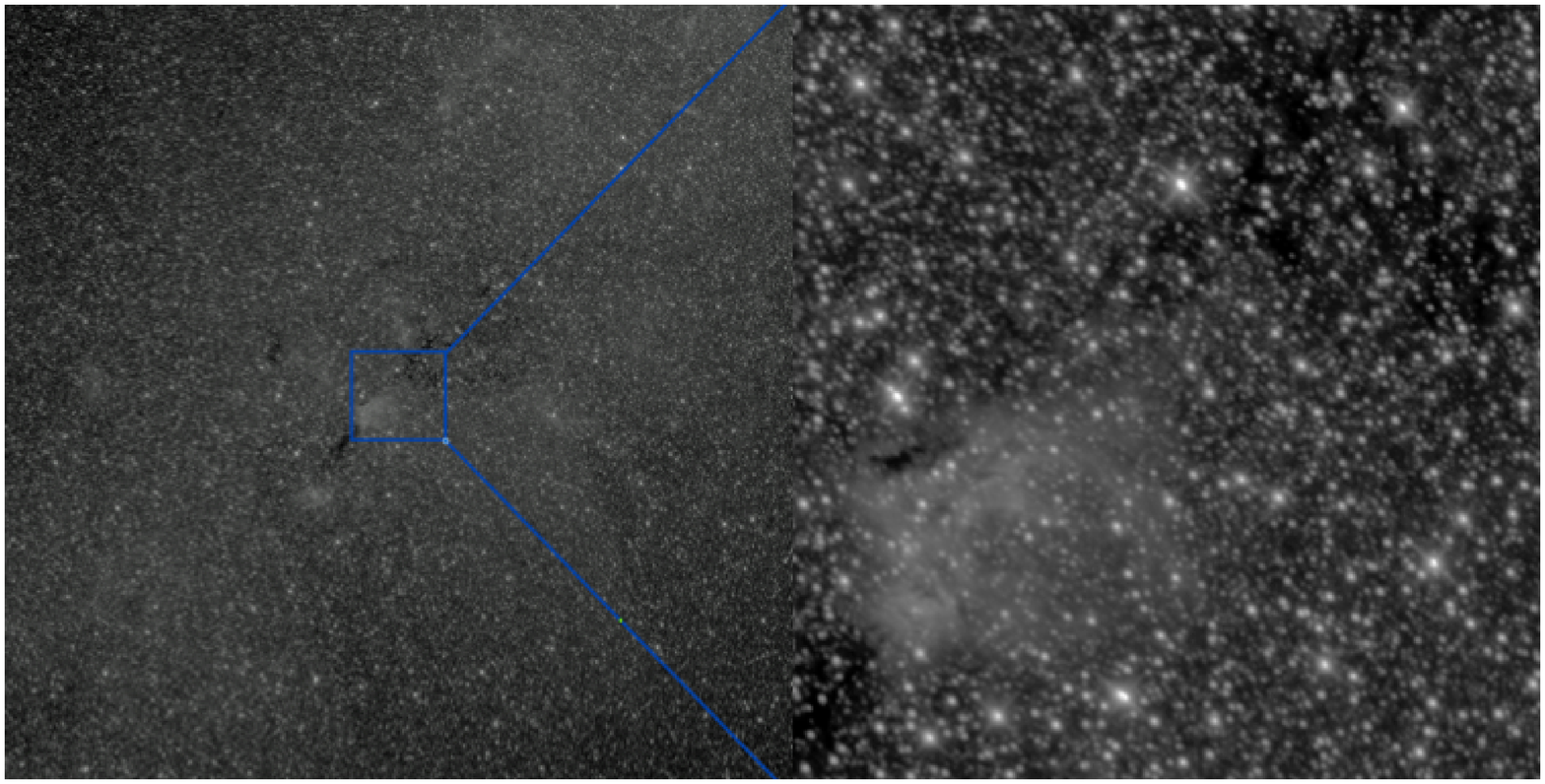}{Figure 2}{ ALLWISE Band 1 (3.4 $\mu$m) production run image.
Left: A single tile of size $4.1\deg \times 4.1 \deg$, shown at HiPS Level 8. Right: Expanded view of central $0.5\deg \times 0.5 \deg$ shown at level 18.}

\end{document}